\documentclass{article}
\usepackage{graphicx} % Required for inserting images

\usepackage[authordate,backend=biber,dashed=false]{biblatex-chicago}
\usepackage{epsfig,endnotes, hyperref, bm}
\usepackage{xcolor}

\usepackage{indentfirst}
\usepackage{breqn}
\usepackage{float}
\usepackage{graphicx,fancyhdr,amsmath,amssymb,amsthm,subfig,xurl,hyperref}
\theoremstyle{definition}

\mathchardef\mhyphen="2D

\usepackage[margin=2cm]{geometry}

\usepackage[utf8]{inputenc}

\title{The Impact of De-Identification on Single-Year-of-Age Counts in the U.S. Census}
\author{Sarah Radway (\textit{Harvard University)}\footnote{\noindent sradway@g.harvard.edu},  
Miranda Christ (\textit{Columbia University})\footnote{\noindent mchrist@cs.columbia.edu}}
\date{ }

\begin{document}

\maketitle

\begin{abstract}

In 2020, the U.S. Census Bureau transitioned from data swapping to differential privacy (DP) in its approach to de-identifying decennial census data. This decision has faced considerable criticism from data users, particularly due to concerns about the accuracy of DP.
We compare the relative impacts of swapping and DP on census data, focusing on the use case of school planning, where single-year-of-age population counts (i.e., the number of four-year-olds in the district) are used to estimate the number of incoming students and make resulting decisions surrounding faculty, classrooms, and funding requests.
We examine these impacts for school districts of varying population sizes and age distributions.

Our findings support the use of DP over swapping for single-year-of-age counts; in particular, concerning behaviors associated with DP (namely, poor behavior for smaller districts) occur with swapping mechanisms as well.
For the school planning use cases we investigate, DP provides comparable, if not improved, accuracy over swapping, while offering other benefits such as improved transparency. 
\newline 
\newline

\end{abstract}

\newpage 

\section{Introduction}\label{sec:intro}

U.S. census data is crucial for education planning, as it's used to determine funding for K-12 schools, colleges, and adult skill development programs \autocite{local_planning, counting_kids, 2020_matters}. 
The U.S. Census Bureau therefore places great focus on school planners when tabulating data, releasing enrollment and spending data for school districts \autocite{school_enrollment, school_enrollment_detailed, student_spending}.
Accurate data allows planners to make estimates which impact funding decisions: to hire teachers, to expand programs and classroom space, or to request additional federal funding.

While the Census Bureau publishes rough total enrollment numbers for school districts, school planners still need to deduce how many students to expect for an incoming preschool or kindergarten class.
A working report by McElrath and Bauman (2021) of the Social Economic and Housing Statistics Division of the Census Bureau estimates preschool enrollment numbers by looking at variation in two different census surveys: ACS and SIPP. 
This report uses Census single-year-of-age data for children age 3 and age 4 to represent children that could be enrolled. 
We will be examining this use case in depth, focusing on how the utility of these types of single-year-of-age based predictions are impacted by de-identification. 

While de-identification mechanisms introduce inaccuracy into census data, they are necessary to protect participant privacy.
Census data contains sensitive information that can be used to create harms against individuals and households: during World War 2, U.S. census data was used to identify Japanese individuals for internment \autocite{internment_camps}.
In addition to its importance in its own right, privacy impacts data accuracy---as noted in Baldrige v. Schapiro, privacy concerns impact census participation rates \autocite{baldrige}, potentially leading to undercounts of populations most concerned about privacy.
Indeed, with more than 53\% of the population being concerned with confidentiality in census data \autocite{abowdtestimony}, a lack of privacy has the potential to significantly impact data utility. 
Moreover, the U.S. Census Bureau is mandated by Title 13 of U.S. Code to de-identify data prior to release \autocite{title13}. 

We compare the accuracy of data de-identified using swapping and differential privacy, focusing on single-year-of-age counts.
We de-identify synthetic data using swapping and differential-privacy--based algorithms and compare the resulting mean absolute percentage error (MAPE).
We study how this error varies by school district size, looking at the accuracy of our mechanisms across small, medium, and large tracts.
We also examine how these de-identification methods affect the accuracy of counts for the total population, the population of age under 18, and the population of ages 4 and 5.
We identify what factors lead certain counties to exhibit worse utility for this use case---namely, population size and age group.

Our findings suggest that both de-identification mechanisms exhibit similar behavior across populations of various sizes, producing higher error for smaller populations and smaller subgroups (i.e. single-year-of-age) than larger populations, and with greater variations in accuracy between parameter values for small populations. 
However, differential privacy allows for greater algorithmic transparency, performs more predictably, and has more similar curves across populations of different sizes.
For school planning data, such as single-year-of-age data, this means that planners should be giving special concern for smaller populations and sub-groups.

\section{Background} \label{sec:background}

Initially, the Census Bureau  de-identified data using combinations of techniques such as rounding or cell suppression \autocite{zayatz2007disclosure}.
These methods entailed generalizing data to a level where it was no longer identifiable. For example, using rounding, instead of reporting that there is one 5-year-old and two 6-year-olds, the tables would instead report 3 children aged 5-6. Or, using cell suppression, the total counts for blocks this small would not be reported--- this is obviously unfavorable for data users, and much of the data suppressed would be that of minorities \autocite{abowdtestimony}. 
Thus, recently, other methods have been introduced which allow for more comprehensive reporting.

In 2000 and 2010 Decennial Censuses, the Census Bureau began to primarily use a disclosure avoidance method known as swapping.
In \textit{swapping}, the algorithm selects data rows from one geographic region and switches them with data rows from another geographic region.
The \textit{swap rate} is a parameter determining how many rows are swapped. 
The Census Bureau's implementation of swapping, which we investigate in greater depth in \autoref{sec:swap}, prioritizes swapping households that are unique within their geographic area.

In the age of large-scale data harvesting, the privacy that data swapping provides is no longer sufficient.
Data brokers seek to learn as much as possible about individuals, using database matching to identify individuals that are present in outside datasets \autocite{federal2014data}.
This weakens the protections provided by some disclosure avoidance mechanisms: data brokers have access to a wealth of information from various websites, phone applications, and more, which can be used to perform database matching with census data.
These concerns were verified in a Census Bureau experiment, where researchers were able to reconstruct swapped data \autocite{abowdreconstruction}, by linking census data to commercial data from companies like Experian Marketing Solutions Incorporated \autocite{abowdtestimony}.
This resulted in as many as 142 million rows being reconstructed \autocite{abowdreconstruction}, which the Census Bureau interpreted as a major concern in light of Title 13.
While the practical relevance of these studies has been contested \autocite{ruggles2022role}, it is clear that database matching poses a threat to public data reporting.

Thus, in order to adhere to the strict privacy standards necessary for a successful census, the Census Bureau adopted differential privacy for the 2020 Census.
\textit{Differential Privacy (DP)} is a mathematical guarantee that ``ensures that only a limited amount of additional risk is incurred by participating in the socially beneficial databases" \autocite{dwork2008differentialsurvey}. 
This guarantee is attained through using mechanisms that add noise to a dataset.
Differential privacy is parameterized by epsilon: a higher epsilon value provides worse privacy and better accuracy, whereas a lower epsilon value provides better privacy and worse accuracy. 
The Census Bureau's implementation of differential privacy, known as the TopDown Algorithm (TDA), adds noise in a method optimized for census use cases, incorporating census-specific invariants (state populations must remain constant) and structural elements, such as households \autocite{wright2020variability, ito_10.1007/978-3-030-57521-2_24}.

\section{Related Work}

Some previous work has examined the level of privacy provided by swapping.
Both Hawes and Rodríguez (2021a) and Ramchandran et al. (2012) investigate the ability to re-identify Census data, with Hawes and Rodríguez (2021) demonstrating that ``low swap rates have essentially no impact on re-identification outcomes", and ``high swap rates have only a minimal impact" \autocite[24]{privacylossbudget}.  
Others have looked at the amount of error introduced by the Census Bureau's swapping algorithm, demonstrating the significant impact on utility caused by prioritizing unique entries when selecting rows to swap \autocite{kim2015effect}.
Ito and Hoshino (2014) examine the usability of swapping for the Japanese census in terms of both accuracy and privacy, proposing 
a combination of targeted and random swapping method that 
selects data rows for swapping both randomly and heuristically
to preserve privacy
\autocite{ito2014data}. 
Similarly, Shlomo et al. (2010) propose a new method of swapping for the UK and US censuses, and compare it to random swapping \autocite{shlomo2010data}.

Several works have examined the impact of DP mechanisms like TopDown Algorithm on data utility.
They focus on its fitness for specific use cases: for example, redistricting use \autocite{cohen2021census, harvardpaper}, public health use \autocite{hauer2021differential, Santos-Lozada13405}, migration tracking \autocite{winkler2022differential}, and allocation of education funds \autocite{steed2022policy}.
Kenny et al. find that after postprocessing has been applied, the TopDown Algorithm offers similar utility to swapping, though it performs poorly for smaller subpopulations and racial minority groups \autocite{kenny2023evaluating}.

Other work draws positive conclusions about DP. 
Cohen et al. (2021) find that DP census data is still fit for use in redistricting. 
Steed et al. (2022) examine the impact of statistical uncertainty on Title 1 funding allocation, demonstrating the impact of differential private mechanisms in comparison with other sources of uncertainty.
They show that the error introduced by DP is small in comparison to these other error sources.
Petti and Flaxman (2019) compare the accuracy and privacy loss of TDA and a simple subsampling privacy method, finding that TDA performs well \autocite{petti2019differential}.

More recent work has taken steps towards measuring the accuracy of both swapping and DP, with respect to the same (synthetic, due to privacy concerns) ground truth dataset. 
Christ et al. (2022) focus on racial minority under-representation, demonstrating that differential privacy provides comparable utility to swapping mechanisms, while providing more desirable privacy protections \autocite{christ2022differential}. 
Similarly, Flaxman and Keyes (2022) examine the ability to identify trans youth using data de-identified using swapping and DP, comparing the potential disclosure associated with each mechanism type \autocite{deidtranskids}.
They conclude that DP offers significantly better privacy than swapping.

The work closest to ours similarly studies the effect of DP on age data released by the Census Bureau \autocite{dyrting2022reconstruction}. 
This work shows that applying a smoothing method, called P-TOPALS, to the output of the DP mechanism yields improved accuracy of age data, including single-year-of-age data. 
They analyze the resulting data's fitness for use in health-focused applications including estimating death rates and fertility rates.
Though their analysis is fairly comprehensive for differential privacy, they do not study swapping.

We will build upon these prior works, in order to provide a comparative analysis of how real world use cases of single-year-of-age data are impacted by both differential privacy and swapping, focusing on the impact of both population size and age group on data utility for real-world use cases surrounding school planning.

\section{Methods}

In order to understand the impact of mechanisms, we first needed (1) mechanisms and (2) data to run the mechanisms on. 
As outlined in Christ et al. (2022), we are unable to use the exact census swapping and differential privacy implementations---details of the Census Bureau's swapping mechanism cannot be released due to privacy concerns, and their DP implementation is too expensive to run for our full suite of analyses.
We develop our data and variants of these mechanisms with the intention that they be representative of those used by the Census Bureau. 

\subsection{Data Generation}

Throughout this work, we examine data at the tract level: aside from being most relevant to the use case of school planning (because tracts are similar in size to school districts), tract-level data is frequently used to develop community support programs, to guide school district decision-making, and to determine many of the federal duties of the Census, which are the functionalities we view as the most important uses of census data \autocite{censustract}.
These census tracts are drawn to range between 1,200 to 8,000 people, or between 480 to 3,200 housing units, with optimal values of 4,000 people or 1,600 housing units \autocite{tractsize}.
Tracts are designated with the intention to align with the needs of the people within them, resulting in the variation in size; they encompass natural geographic boundaries, and take into account factors such as American Indian reservations \autocite{tractsize}.

Because the true microdata collected by the Census Bureau is not publicly available, we use synthetic microdata that we treat as the ground truth throughout our experiments.
Our microdata was generated using a method of Flaxman and Haddock; the associated code of which is publicly available \autocite{flaxmanrepo}. 
This library uses integer programming to reconstruct microdata matching the tables in Summary File 1 of the Census Bureau's published 2010 Decennial Census Data.
The resulting synthetic data thus closely matches the characteristics of the populations they simulate.

\begin{figure}[!htbp] 
\begin{center}
\caption{\label{fig:tracts} An overview of the synthetic tracts used for our analysis.}
\begin{tabular}{|c|c|}
\hline
Tract Name & Population Count  \\
\hline 
RI-40200 & 1,158 \\
\hline
RI-51000 & 1,508 \\
\hline
RI-18000 & 2,632 \\
\hline
RI-15800 & 3,771 \\ 
\hline
AL-5400 & 4,068 \\
\hline
RI-20101 & 8,415 \\
\hline
AL-100 & 12,267 \\
\hline
\end{tabular}
\end{center}
\end{figure} 

We use synthetic state data for various tracts in Rhode Island and Alabama. 
To simulate school districts of varying sizes, 
we select seven tracts ranging in population from 1,158 to 12,267.
The full selection of tracts is attached in \autoref{fig:tracts}. 
We use our ground truth synthetic data to create two sets of Privacy-Protecting Microdata Files (PPMFs): one using swapping and one using differential privacy.

\subsection{Swapping Implementation} \label{sec:swap}

Swapping is a disclosure avoidance method that involves exchanging data from one dataset with data from another dataset. 
Since the publication of Christ et al., a recent paper from Abowd and Hawes has provided a much clearer view of the Census Bureau's swapping mechanism \autocite{abowd2023confidentiality}.
Here, we make slight modifications to the mechanism used in Christ et al. to more accurately capture the Census Bureau's mechanism.

In the case of the U.S. Decennial Census, swapping occurs at the household level \autocite{abowd2023confidentiality}. 
This means that, if selected for swapping, a household's data would be exchanged with data about another household within the same state.
The Census Bureau's implementation prioritizes selecting households that are identifiable (unique) within their geographical areas \autocite{gordonlong}.
Therefore, we iterated over each household, and identified households of unique combinations of household member age, sex, and race. 
We prioritize these households to be selected for swapping.

As outlined in a recent publication from the Census Bureau, when choosing what household to swap with the selected household, the mechanism prioritizes two features: maintaining household size, and maintaining the number of household members over and under 18 years of age (of non-voting age) \autocite{abowd2023confidentiality}. 
Therefore, for each household selected for swapping, we randomly select a household from the dataset of the entire state that matches for these two variables.

Our implementation, and that of the Census Bureau, is parameterized by the \textbf{swap rate}, which is the proportion of households that are swapped.
We choose our swap rates used in our experiments based on published U.S. Census Bureau guidance \autocite{census-sr-value}; which states that up to half of the households in a block are swapped (0-50\%). 
While the exact swap rates used for each geographic area cannot be released, as this would facilitate re-identification \autocite{mervis2019can}, conversations suggest that the national estimated swap rate is the range of 2-4\%, varying between regions. 

\subsection{DP Implementation}

As with Christ et al. (2022), financial limitations render us unable to run the true TDA mechanism; we thus implement the same alternative implementation of differential privacy (DP), meant to capture the ``general behavior of TDA" [462]\autocite{christ2022differential}.
While a full overview of their mechanism, and its resemblance to TDA, can be found in the original work, the main components of this mechanism are that it uses the geometric mechanism, which is parameterized by $\epsilon$, from IBM's Differential Privacy Library \autocite{holohan2019diffprivlib}.
In this work, we add noise by first computing a single-year-of-age histogram over the reconstructed microdata file, and add independent noise according to our geometric mechanism to each of these age counts.
We perform no post-processing, and we note that the accuracy of our DP data may be further improved by using a smoothing method as suggested in \autocite{dyrting2022reconstruction}; our mechanism will exhibit strictly worse accuracy than TDA.

\subsection{Parameter Selection}

We select our parameter ranges (swap rates and epsilon values) to look at reasonable ranges for our specific evaluations.

For swapping, though the national average likely rests between 2-4\%, for groups with greater subgroup diversity, this swap rate will likely be much higher, to account for the greater number of unique entries.
Thus, we generally can be looking in the .02-.04 range, especially for larger tracts.
For smaller tracts, there will likely be more unique entries, so our intuition tells us that the swap rate will be higher.

In the case of differential privacy, we plot the error for epsilon in the range of 0-10, to ensure all reasonable epsilon values are included for the selected tracts. 
These epsilon values and swap rates shown are meant to capture a range of reasonable values, we do not intend to provide direct correlations between swap rates and epsilon values. 

\subsection{Accuracy Metric}

Because our use case for school planning is focused on the accuracy of single-year-of-age counts, we compute histograms of these age counts for each of our dataset types: the ground truth data, the swapped data, and the differentially private data. 
For each of our de-identified datasets, we compute the Mean Absolute Percentage Error (MAPE) between its histogram and the ground truth histogram. The MAPE is an accuracy metric commonly used by the Census Bureau \autocite{datametrics, hawesspence}.
Recall that we analyze the error for several age groups: the total population, the population under 18, and the population aged 4-5.
We first restrict the unmodified histogram and the modified histogram to the range in question, then for each age, we compute the absolute error and divide it by the true count of that age group from the modified to obtain the absolute percentage error. 
If this true count is zero, we omit it. 
We compute the sum of the non-omitted absolute percentage errors of our selected ages (e.g., ages 4 and 5), then divide by the number of selected ages (e.g., 2) to obtain the average.
For each swap rate and epsilon value, we computed the MAPE over five separate swapping and differential privacy runs.

\section{Results}

As we begin to evaluate utility for a data user, we briefly focus on the total population (including all ages), then examine data more relevant for school planning use cases: the accuracy of counts for the population ages under 18, and accuracy of counts for the population ages 4-5. 
In this way, we can interpret the impact of our de-identification mechanisms on data utility, and how factors such as population size and age group alter this impact. 

Our figures show the epsilon rate on the bottom x-axis, the swap rate on the top x-axis, and the MAPE on the y-axis.
Privacy decreases from left to right.
Note that vertically aligned swap rates and epsilon values are not necessarily equivalent. 
While the individual values don't match up exactly, the overall ranges of epsilon and the swap rate that we plot reflect those used by the Census Bureau. 

\newpage

\subsection{Total Population}

\begin{figure}[!htbp]
    \begin{center}
    \includegraphics[width = 0.6\textwidth]{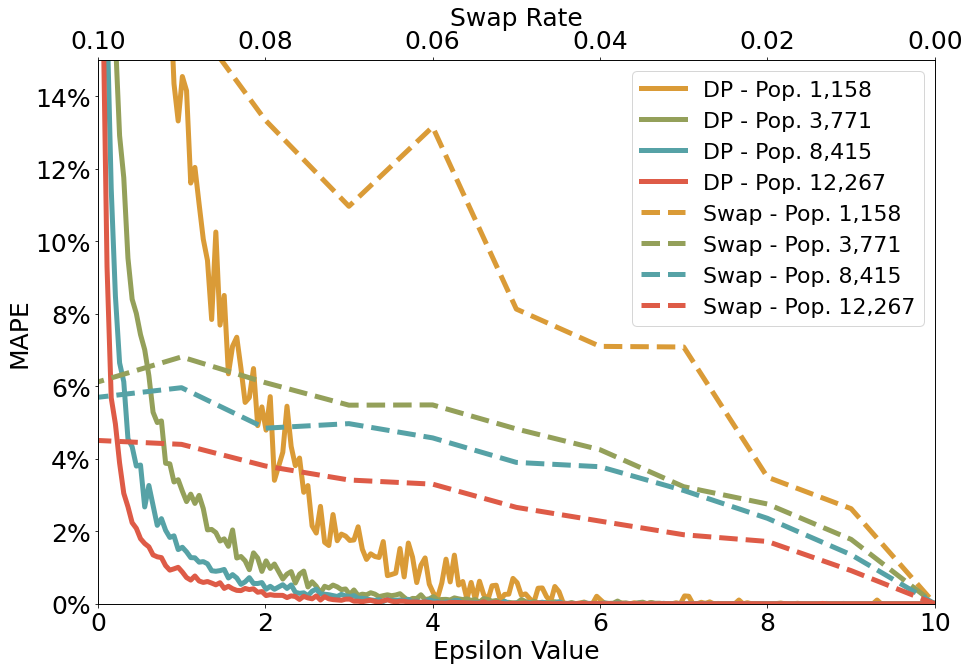}
    \caption{\label{fig:total} The average MAPE of single-year-of-age data over the total population, for data produced using DP and swapping for tracts of varying population sizes.} 
    \end{center}
\end{figure}

%how to interpret figs
In \autoref{fig:total}, we see the MAPE for four tracts of varying size, ranging from roughly 1,000 to 12,000 (RI-40200, RI-15800, RI-20101, and AL-100). 

%mape & pop size: smaller populations = worse mape
There is a clear relationship between MAPE and population size, regardless of de-identification mechanism: larger populations have lower error, and smaller populations have higher error. 
Across the three larger tracts, the difference in MAPE is not dramatic: a mechanism produces similar results (somewhere within a 2\% range) for each of these tracts for any given parameter value. 
For the smallest tract, with a population of 1,158, we see that the MAPE is significantly higher than the three larger tracts, for both differential privacy and swapping; as tract sizes get smaller, MAPE quickly increases. 

% something about how error is highest for total pop
Compared to narrower age groups, the total population has the highest error across the groups we examine. 
This is largely because it includes individuals of ages over 75, which generally have low population counts. 
The relative error added by DP is higher for these smaller groups.
%trends consistent across pop sizes
Across the various tracts, the trends for the respective de-identification mechanisms remain consistent: differential privacy roughly follows a decaying exponential, whereas swapping has a slightly unpredictable, but linear-like decay. 
We also observe that the variation in error between parameter values for a given mechanism becomes greater for smaller tract sizes. 
We see that differential privacy is capable of providing comparable or improved MAPE values to those of swapping within the selected range of parameter values.

\subsection{Under 18}

\begin{figure}[!htbp]
    \begin{center}
    \includegraphics[width = 0.6\textwidth]{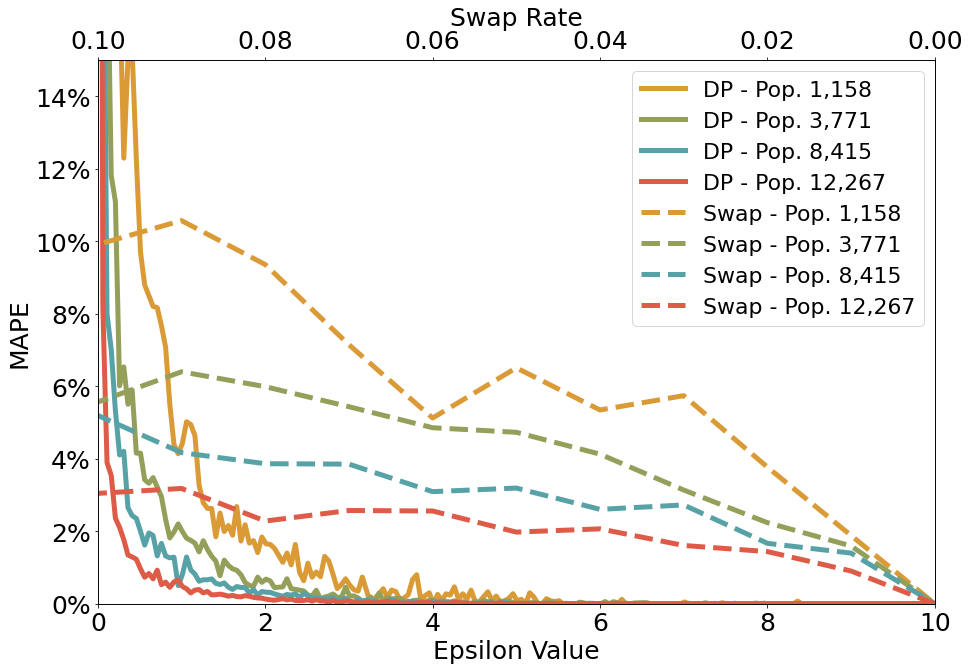}
    \caption{\label{fig:eighteen} The average MAPE of single-year-of-age data over the population under 18 years of age, for data produced using DP and swapping for tracts of varying population sizes} 
    \end{center}
\end{figure}

%under 18 use case
Next, we limit our analysis to looking only at MAPE for the population under 18 years of age. 
This value could, for example, provide a general idea of the total number of students that could potentially enroll in daycare, preschool, and school.

%swapping
This age limitation is particularly notable for swapping; because Decennial Census data is used to ``redraw congressional, state, and local district boundaries'' \autocite{apportionment}, it is particularly important to preserve the utility of counts of residents of voting age vs. residents of non-voting age.
For this reason, the swapping mechanism requires that households that are swapped with each other must have the same number of members above and below eighteen. 
We do not see or expect a value of 0.0: while the count of individuals under 18 is invariant under swapping, and thus equal to 0, the count of 1-year-olds or 17-year-olds, for example, is not invariant. 
There will therefore instances where a 1-year-old is swapped with a 17-year-old, introducing inaccuracy for single-year-of-age counts.
However, we would expect, and observe, that swapping performs slightly better for the population under 18 in comparison with the total population. 
This is because we expect a greater level of data preservation for this age group---for an individual under 18, there are only 17 other values to be swapped with.

For differential privacy, we observe slightly improved results compared to those for DP over the total population.
This is because there are many ages over 75 years with very low population counts, contributing error in the total population but not in the population under 18. 
Across our tracts, we see similar error values to those of the total population, with, once again, differential privacy producing better accuracy than swapping in our range of parameter values. 

\subsection{Ages 4-5}

\begin{figure}[!htbp]
    \begin{center}
    \includegraphics[width = 0.6\textwidth]{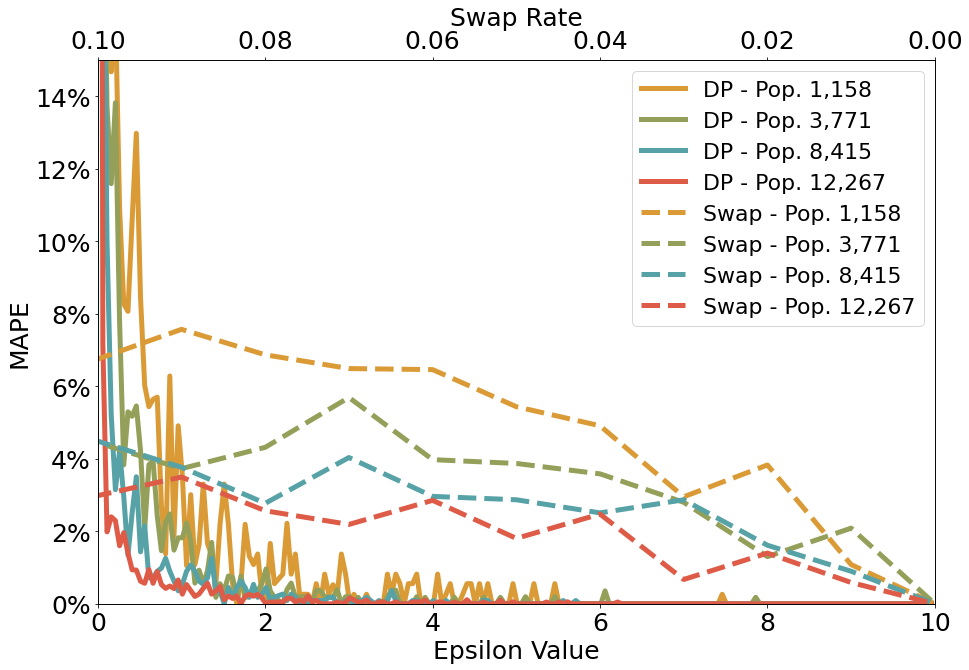}
    \caption{\label{fig:kind} The average MAPE of single-year-of-age data over the population aged 4 and 5, for data produced using DP and swapping for tracts of varying population sizes} 
    \end{center}
\end{figure}

%use case for ages 4-5
Lastly, we consider a use case that the swapping mechanism's implementation is less explicitly designed to accommodate.
One significant school planning use case for Decennial Census data is estimating the number of incoming kindergarteners a school should expect.
These estimates are crucial to ensure that proper decisions are made surrounding classroom space, teaching staff, and funding requests.
The accuracy of single-year-of-age data is crucial here--knowing the number of children of kindergarten age helps a school planner to understand the number of enrollments to expect.
We examine the utility for this use case, looking at the MAPE for the population subset of ages 4-5. 

%trends present
Generally, we see the same trends, but with greater variation between parameter values: as shown in \autoref{fig:kind}, both differential privacy and swapping show more significant variations in MAPE between epsilon values, and between swap rates. 
This is because the age group (4-5) is narrower (compared to the total population): in general, we see greater variation for smaller population segments for both differential privacy and swapping.

\subsection{Examining an Outlier Tract}

\begin{figure}[!htbp]
    \begin{center}
    \includegraphics[width = 0.6\textwidth]{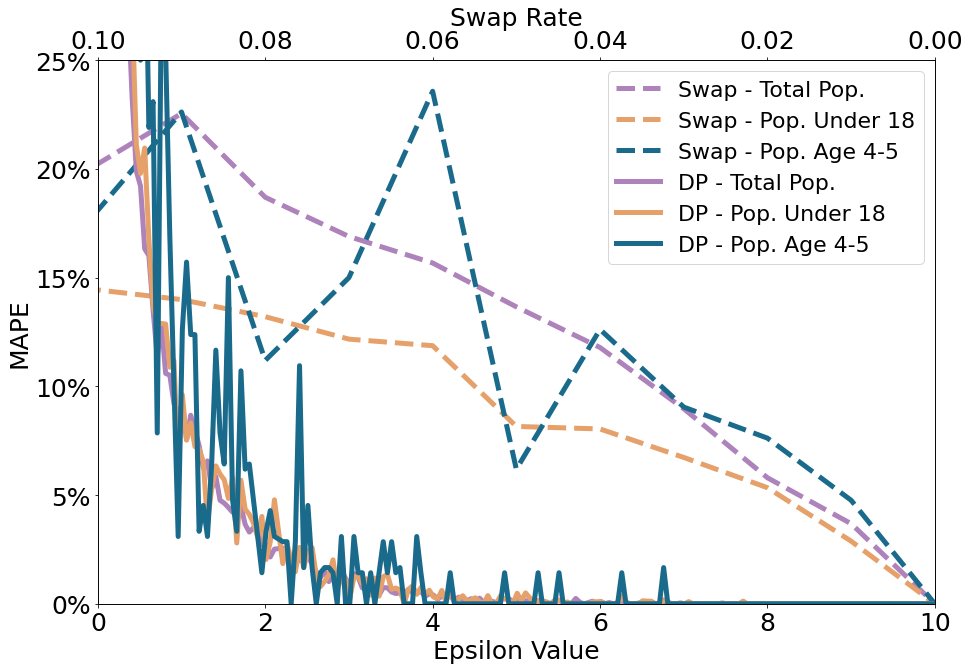}
    \caption{\label{fig:51000} MAPE For single-year-of-age data; comparing total population, population age under 18, and population age 4-5. Shown for differentially private and swapping mechanisms.} 
    \end{center}
\end{figure}

We also sought to identify the impact of de-identification on tracts that exhibit unique qualities, causing them to be outliers in terms of performance. 
We focus on a tract, RI-51000, which exhibits qualities that differ from a typical tract: namely, this county has a population of 1,508, where a large chunk of this tract consists of nursing homes.
This tract is thus especially atypical in its age distribution--for example, in RI-51000 there are only 13 kindergarteners in the entire tract, as compared with RI-42000, a tract with a significantly smaller population of 1,158, which has 56 kindergartners.

Through this example, we can examine how small population size, and particularly how this unusual age distribution, impacts data utility.
Looking at the same three population groupings as above, we see the trends are present, but more defined: with higher error for smaller populations, and with especially higher and less smooth MAPE for age 4-5, which is limited in representation.
Because this total population (and population under 18) is smaller, we see significantly increased MAPE compared with other, larger tracts.
This falls in line with our findings from earlier tracts: as the total population size shrinks, the utility after de-identification will also lessen. 
Across the board, we see the MAPE is more than twice that of RI-15800, demonstrating that MAPE significantly increases as population size decreases, particularly for narrower age groups.

However, we particularly see significant variations in performance for the kindergartner age group: the range in MAPE across parameter values varies dramatically: this is due to the incredibly small size of the dataset.
As expected, we see that noise added through our differentially private mechanism causes more significant impact on data utility and consistency---however, we see that our swapping mechanism demonstrates these same behaviors, also producing dramatically different results for differing parameter values. 

\subsection{Takeaways}

When we investigate how swapping and differentially private mechanisms behave across our various tracts, there are a few trends that emerge.

We see that the utility of both mechanisms is lower for smaller populations. 
For swapping and differential privacy, larger populations have lower error, and smoother, more predictable curves. 
Alternatively, for smaller populations the mechanisms have higher error, and behave less predictably.
This is represented by the high MAPE values, and jagged nature of the MAPE between various parameter values, particularly in Figures \ref{fig:51000} and \ref{fig:kind}.
For tracts that are larger, and with more balanced sub-groups, the MAPE of swapped microdata is more consistent.
This unpredictability is especially dramatic with swapping, because the algorithm itself is expected to run in an unpredictable manner \autocite{christ2022differential}.

For differential privacy, this error is higher for smaller groups simply because the algorithm is expected to run on the data independently; noise is added consistently across various populations. 
In larger populations, where the number of individuals of each age is greater, the proportion of error to true age count is smaller.
As we shrink population size, differential privacy largely shows the same trend as swapping, with both higher variation and higher error in the smaller tracts. 
Thus, we see that both of these de-identification mechanisms had greater negative impact on data utility when run on smaller populations. 

This inverse relationship between the percentage error incurred by differential privacy and the population count that noise is being added to also appears in the MAPE of the three age subgroups that we examine. 
The MAPE of differential privacy decreases slightly from the total population, to the population under 18, to the population aged 4-5, as shown in Figures \ref{fig:total}, \ref{fig:eighteen}, and \ref{fig:kind} respectively.
The error is highest for the total population because in our tracts, there are few residents of the higher ages (75+) compared to the lower ages, with very few residents older than 90. 
Relative to these low counts, differential privacy adds substantial noise, increasing the error for the total population.
The under 18 subset and aged 4-5 subset do not include these higher ages with very low population counts, resulting in lower error. 
The error for under 18 and ages 4-5 is similar, with ages 4-5 having slightly higher population counts and slightly lower error.

We also see that the utility of age groups is impacted by the size of the age group (i.e., the number of kindergarten age children). 
This was demonstrated by looking at an outlier county with a large elderly population, RI-51000: we see that small age groups, even in larger tracts, will be most dramatically impacted by de-identification mechanisms.

In our investigation, the attribute evaluated (age) is directly involved in the swapping mechanism: households are swapped based upon the number of household members over and under 18. 
This allows for some level of correlation to seep through into the results.
In our implementation, performance for swapping will be worse for an attribute where this is not the case.
This is not a consideration for differential privacy. 

Across our tracts, even when performance is significantly degraded (ex. \autoref{fig:51000}) it appears that differential privacy consistently performs about as well, if not better, than swapping at reasonable parameter values.

\section{Discussion}

\textbf{Small groups are particularly negatively impacted by both de-identification mechanisms}. In terms of impact on school planning (and other similar use cases), data users have to be careful accounting for error in these small groups. 
We have seen that single-year-of-age data can be have significant error, particularly for small tracts. 
Consideration will be needed when making decisions based upon this data---clearly, it should not be treated as ground truth.

However, this is not just the case for differential privacy. 
When it comes to comparatively evaluating swapping and differential privacy, differential privacy is comparable, if not an improvement upon, swapping for all reasonable parameter values at the tract level. 
We see that many of the criticisms of differential privacy are also behaviors present in swapping mechanisms as well. 
This provides justification for the switch to differential privacy, in the context of this use case.

\textbf{Differential privacy allows for transparency in de-identification}. While de-identification has always occurred, differential privacy is the first time that the public has been able to understand and quantify the impact of disclosure avoidance mechanisms' impact on accuracy. 
While differential privacy allows for the publication of epsilon values, previous methods like swapping could not reveal their parameter values without risking identifying the ground truth data. 
Similarly, differential privacy allows for a far greater range of data publication: previously, it was required that ``all categorical variables must have at least 10,000 people nationwide in each published category'', or else they would be recoded \autocite[6]{zayatz2007disclosure}. 
Systems like TDA can lift this requirement.

\textbf{Transparency is necessary for stronger decision making}. 
As data users begin to grapple with the full impact of past de-identification on data utility, and as future mechanisms are developed, it becomes clear that transparency is a necessary aspect of de-identification: it is necessary to understand the extent to which these mechanisms are impacting data accuracy.

Differential privacy lends itself well to this requirement: it allows for publication of epsilon values, helping us understand the amount of inaccuracy introduced without negating its benefits. 
Differential privacy gives data users the ability to make better decisions based off the modified data, through providing an idea of how far the reported value might be from the true value. 
For example, if a school planner sees a reported value of seven kindergarteners (a small group) with a high epsilon value, then they know there's probably a lot of error that needs to be accounted for; they can choose to plan for ten students, rather than seven, or perhaps realize that they shouldn't be making decisions based off of this data at all. 
It provides a glimpse into the trustworthiness of the data.
For swapping, there is not the same context surrounding how inaccurate the data might be, since we can't be transparent with the swap rate. 
That means data users have to make resulting decisions based only off of reported total population or sub population size, without the insight into the level of data modification provided by a parameter value. 

Because differential privacy can provide these traits, while providing similar levels of utility to swapping, our results suggest differential privacy is a better option for de-identification of tract data for public use cases such as school planning.

\section{Limitations and Further Work}

Our study is limited by both computational capability and data availability. 
Our work did not use the Census Bureau's TDA, as the cost would have proven infeasible for our work. 
Cost also limited the number of tracts that we were able to run for our analysis: if we were able to run more tracts, we would be able to draw more definitive conclusions.
Further, we do not have access to ground truth data: only the U.S. Census Bureau knows true tract parameter values.
Thus, our mechanisms were running on synthetic data, which possibly could induce small differences in behavior. 
Lastly, we do not know the true parameter values used by the Census Bureau. 
While we can look at the range of likely values, we are unable to provide a direct comparison between a swap rate and a epsilon value selected by the Census Bureau.  

While we believe that our design choices are grounded in publicly available Census Bureau guidance, and should be generally representative of the true data and mechanisms, we suggest that a larger scale study would prove useful and fruitful conclusions. 
We believe if the Census Bureau were to run similar experiments, with true ground truth data, and the actual parameter values they selected, this could provide compelling argument regarding de-identification methods, through providing some limited, real-world comparison to help demonstrate the impact of de-identification mechanisms, and help us understand the reasoning behind their choices of mechanisms.
However, we acknowledge that this is unlikely, as the Census Bureau would likely have reservations on the grounds of privacy concerns.

Further work, which we hope to undertake, could investigate how to communicate with data users about the impact of these mechanisms; now that the Census Bureau is able to have greater transparency surrounding data de-identification, it is useful to understand how data users make decisions with de-identified data, and their consideration for the impact of de-identification.
We hope to explore these questions moving forward.

\section{Conclusion} \label{sec:conclusion}

In the case of single-year-of-age data, our work suggests that de-identification mechanisms inherently impact utility in seemingly similar ways. 
Both mechanisms have higher error for smaller populations and small sub-populations: this merits consideration as a data user.  
For swapping, the accuracy was more unpredictable between swap rates, especially in these smaller tract populations. 
Differential privacy performed somewhat more predictably across different tract sizes, but still decreased utility as population size shrinks. 
However, we find the properties of differential privacy a bit more desirable: namely, for the ability to provide transparency surrounding mechanism performance. 
Thus, in the school planning use case, and beyond, we demonstrate the merit of differential privacy as a de-identification mechanism.

\subsubsection*{Acknowledgments:} 
We thank
    Steven M. Bellovin,
    danah boyd,
    Abraham Flaxman,
    and Johes Bater
    for their assistance and many helpful comments.

\printbibliography

@article{kenny2023evaluating,
  title={Evaluating Bias and Noise Induced by the US Census Bureau's Privacy Protection Methods},
  author={Kenny, Christopher T and Kuriwaki, Shiro and McCartan, Cory and Simko, Tyler and Imai, Kosuke},
  journal={arXiv preprint arXiv:2306.07521},
  year={2023}
}

@misc{title13,
  title={13 {U.S. Code} § 9},
  howpublished={Available at: \url{https://www.law.cornell.edu/uscode/text/13/9} (Accessed: May 2022)},
}

@misc{abowdtestimony,
    title={ 2010 Declaration of John Abowd, State of Alabama v. United States Department of Commerce. Case No. 3:21-CV-211-RAH-ECM-KCN.},
    author={John Abowd},
    howpublished={Available at: \url{https://vhdshf2oms2wcnsvk7sdv3so.blob.core.windows.net/thearp-media/documents/Declaration_of_John_M._Abowd.pdf} (Accessed: August 2022)},
    year={2021}
}

@misc{local_planning,
    title={How Census Data Leads to Local Planning and Funding},
    author={Jacqueline Byers},
    publisher ={American Bar Associaton},
    year={2020},
    % month={May},
    note={Available at: \url{https://www.americanbar.org/groups/public_education/publications/insights-on-law-and-society/volume-20/issue-2/how-census-data-leads-to-local-planning-and-funding/} (Accessed: April 2023)}
}

@misc{counting_kids,
    title={Counting All Kids: How the Census Impacts Education},
    author={Tiffany McDole and Emily Brixey},
    publisher ={EdNote},
    year={2020},
    % month={May},
    note={Available at: \url{https://ednote.ecs.org/counting-all-kids-how-the-census-impacts-education/} (Accessed: August 2022)}
}

@article{internment_camps, 
 title={Confirmed: The {U.S. Census Bureau} Gave Up Names of {Japanese-Americans} in {WW II}}, 
 journal={Scientific American}, 
 publisher={Scientific American}, 
 author={Minkel, JR}, 
 year={2007}, 
  note={Available at: \url{https://www.scientificamerican.com/article/confirmed-the-us-census-b/} (Accessed: August 2022)}
 % month={March}
}

@misc{student_spending,
    title={{Public School Spending Per Pupil Increases by Largest Amount in 11 Years}},
    author={{U.S. Census Bureau}},
    publisher ={U.S. Census Bureau},
    year={2021},
    % month={May},
    note={Available at: \url{https://www.census.gov/newsroom/press-releases/2021/public-school-spending-per-pupil.html} (Accessed: August 2022)}
}

@misc{2020_matters,
    title={{Why 2020 Matters for Schools}},
    author={{U.S. Census Bureau}},
    publisher ={U.S. Census Bureau},
    year={2021},
    % month={October},
    note={Available at: \url{https://www.census.gov/programs-surveys/sis/2020census/why-2020-matters.html} (Accessed: August 2022)}
}

@misc{school_enrollment,
    title={{School Enrollment}},
    author={{U.S. Census Bureau}},
    publisher ={U.S. Census Bureau},
    year={2022},
    % month={July},
    note={Available at: \url{https://www.census.gov/topics/education/school-enrollment.html} (Accessed: August 2022)}
}

@misc{school_enrollment_detailed,
    title={{School Enrollment in the United States: October 2020 - Detailed Tables}},
    author={{U.S. Census Bureau}},
    publisher ={U.S. Census Bureau},
    year={2022},
    % month={July},
    note={Available at: \url{https://www.census.gov/topics/education/school-enrollment.html} (Accessed: August 2022)}
}

@article{zayatz2007disclosure,
  title={Disclosure avoidance practices and research at the US Census Bureau: An update},
  author={Zayatz, Laura},
  journal={Journal of Official Statistics},
  volume={23},
  number={2},
  pages={253},
  year={2007},
  publisher={Statistics Sweden (SCB)}
  % doi={}
}

@article{federal2014data,
  title={Data brokers: A call for transparency and accountability},
  author={{Federal Trade Commission}},
  journal={Washington, DC},
  year={2014},
  note={Available at: \url{https://www.ftc.gov/reports/data-brokers-call-transparency-accountability-report-federal-trade-commission-may-2014} (Accessed: August 2022)}
}

@inproceedings{abowdreconstruction,
  title={Staring-down the database reconstruction theorem},
  author={Abowd, John M},
  booktitle={Joint Statistical Meetings, Vancouver, BC},
  pages={234},
  year={2018},
  note={Available at: \url{http://www.census.gov/content/dam/Census/newsroom/press-kits/2018/jsm/jsm-presentation-database-reconstruction.pdf} (Accessed: August 2022)}
}

@article{ruggles2022role,
  title={The role of chance in the census bureau database reconstruction experiment},
  author={Ruggles, Steven and Van Riper, David},
  journal={Population Research and Policy Review},
  volume={41},
  number={3},
  pages={781--788},
  year={2022},
  publisher={Springer}
}

@inproceedings{dwork2008differentialsurvey,
  title={Differential privacy: A survey of results},
  author={Dwork, Cynthia},
  booktitle={Theory and Applications of Models of Computation: 5th International Conference, TAMC 2008, Xi’an, China, April 25-29, 2008. Proceedings 5},
  pages={1--19},
  year={2008},
  organization={Springer}
}

@article{wright2020variability,
  title={Variability Assessment of Data Treated by the TopDown Algorithm for Redistricting},
  author={Wright, Tommy and Irimata, Kyle},
  journal={Statistics},
  pages={02},
  year={2020},
}

@InProceedings{ito_10.1007/978-3-030-57521-2_24,
author="Ito, Shinsuke
and Miura, Takayuki
and Akatsuka, Hiroto
and Terada, Masayuki",
title="Differential Privacy and Its Applicability for Official Statistics in {Japan} -- A Comparative Study Using Small Area Data from the {Japanese} Population Census",
booktitle="Privacy in Statistical Databases",
year="2020",
publisher="Springer International Publishing",
pages="337--352",
}

@misc{privacylossbudget,
    title={Determining the Privacy-loss Budget},
    author={Hawes, Michael and Rodríguez, Rolando A.},
    publisher={U.S. Census Bureau},
    year={2021},
    howpublished={Available at: \url{https://www2.census.gov/about/partners/cac/sac/meetings/2021-05/presentation-research-on-alternatives-to-differential-privacy.pdf} (Accessed Aguust 2022)}
}

@article{kim2015effect,
  title={The Effect of Data Swapping on Analyses of {American Community Survey} Data},
  author={Kim, Nicolas},
  journal={Journal of Privacy and Confidentiality},
  volume={7},
  number={1},
  year={2015}
}

@inproceedings{ito2014data,
  title={Data swapping as a more efficient tool to create anonymized census microdata in Japan},
  author={Ito, Shinsuke and Hoshino, Naomi},
  booktitle={Privacy in Statistical Databases},
  pages={1--14},
  year={2014}
}

@inproceedings{shlomo2010data,
  title={Data swapping for protecting census tables},
  author={Shlomo, Natalie and Tudor, Caroline and Groom, Paul},
  booktitle={International Conference on Privacy in Statistical Databases},
  pages={41--51},
  year={2010},
  organization={Springer}
}

@inproceedings{cohen2021census,
  title={Census TopDown: The Impacts of Differential Privacy on Redistricting},
  author={Cohen, Aloni and Duchin, Moon and Matthews, JN and Suwal, Bhushan},
  booktitle={2nd Symposium on Foundations of Responsible Computing},
  year={2021}
}

@article{harvardpaper,
  title={The impact of the {US Census} Disclosure Avoidance System on redistricting and voting rights analysis},
  author={Kenny, Christopher T and Kuriwaki, Shiro and McCartan, Cory and Rosenman, Evan and Simko, Tyler and Imai, Kosuke},
  journal={arXiv preprint arXiv:2105.14197},
  year={2021}
}

@article{hauer2021differential,
  title={Differential privacy in the 2020 Census will distort {COVID-19} rates},
  author={Hauer, Mathew E and Santos-Lozada, Alexis R},
  journal={Socius},
  volume={7},
  pages={2378023121994014},
  year={2021},
  publisher={SAGE Publications}
}

@article {Santos-Lozada13405,
	author = {Santos-Lozada, Alexis R. and Howard, Jeffrey T. and Verdery, Ashton M.},
	title = {How differential privacy will affect our understanding of health disparities in the United States},
	volume = {117},
	number = {24},
	pages = {13405--13412},
	year = {2020},
	doi = {10.1073/pnas.2003714117},
	publisher = {National Academy of Sciences},
	issn = {0027-8424},
	URL = {https://www.pnas.org/content/117/24/13405},
	journal = {Proceedings of the National Academy of Sciences}
}

@article{winkler2022differential,
  title={Differential privacy and the accuracy of county-level net migration estimates},
  author={Winkler, Richelle L and Butler, Jaclyn L and Curtis, Katherine J and Egan-Robertson, David},
  journal={Population Research and Policy Review},
  volume={41},
  number={2},
  pages={417--435},
  year={2022},
  publisher={Springer}
}

@article{steed2022policy,
  title={Policy impacts of statistical uncertainty and privacy},
  author={Steed, Ryan and Liu, Terrance and Wu, Zhiwei Steven and Acquisti, Alessandro},
  journal={Science},
  volume={377},
  number={6609},
  pages={928--931},
  year={2022},
  publisher={American Association for the Advancement of Science}
}

@article{petti2019differential,
  title={Differential privacy in the 2020 US census: what will it do? Quantifying the accuracy/privacy tradeoff},
  author={Petti, Samantha and Flaxman, Abraham},
  journal={Gates open research},
  volume={3},
  year={2019},
  publisher={Gates Foundation-Open Access}
}

@inproceedings{christ2022differential,
  title={Differential Privacy and Swapping: Examining De-Identification's Impact on Minority Representation and Privacy Preservation in the US Census},
  author={Christ, Miranda and Radway, Sarah and Bellovin, Steven M},
  booktitle={2022 IEEE Symposium on Security and Privacy (SP)},
  pages={1564--1564},
  year={2022},
  organization={IEEE Computer Society}
}

@misc{deidtranskids,
    title={How Census Data Put Trans Children at Risk},
    author={ Os Keyes and Abraham D. Flaxman},
    publisher={Scientific American},
    year={2022},
    month={September},
    note={Available at: \url{https://www.scientificamerican.com/article/how-census-data-put-trans-children-at-risk/} (Accessed: 2022)}
}

@article{dyrting2022reconstruction,
  title={Reconstruction of age distributions from differentially private census data},
  author={Dyrting, Sigurd and Flaxman, Abraham and Sharygin, Ethan},
  journal={Population Research and Policy Review},
  volume={41},
  number={6},
  pages={2311--2329},
  year={2022},
  publisher={Springer}
}

@article{abowd2023confidentiality,
  title={Confidentiality protection in the 2020 US Census of population and housing},
  author={Abowd, John M and Hawes, Michael B},
  journal={Annual Review of Statistics and Its Application},
  volume={10},
  pages={119--144},
  year={2023},
  publisher={Annual Reviews},
  doi={10.1146/annurev-statistics-010422- 034226}
}

@misc{gordonlong,
    title={Formal Privacy Methods for the 2020 Census},
    author={Gordon Long},
    publisher = {U.S. Census Bureau},
    year={2020},
    note={Available at: \url{https://www2.census.gov/programs-surveys/decennial/2020/program-management/planning-docs/privacy-methods-2020-census.pdf} (Accessed August 2022)}
}

@misc{census-sr-value,
  title={Determining the Privacy-loss Budget: Research into Alternatives to Differential Privacy},
  author={Michael Hawes and Rolando A. Rodríguez},
  journal={Census Scientific Advisory Committee},
  publisher={U.S. Census Bureau},
  HowPublished={Available at: \url{https://www2.census.gov/about/partners/cac/sac/meetings/2021-05/presentation-research-on-alternatives-to-differential-privacy.pdf} (Accessed August 2022)},
  year={2021}
}

@article{mervis2019can,
  title={Can a set of equations keep US census data private},
  author={Mervis, Jeffrey},
  journal={Science},
  volume={10},
  year={2019},
  HowPublished={Available at: \url{https://www.science.org/content/article/can-set-equations-keep-us-census-data-private} (Accessed August 2022)}
}

@article{holohan2019diffprivlib,
  title={Diffprivlib: the {IBM} differential privacy library},
  author={Holohan, Naoise and Braghin, Stefano and Mac Aonghusa, P{\'o}l and Levacher, Killian},
  journal={arXiv preprint arXiv:1907.02444},
  year={2019}
}

@misc{datametrics,
    title={Data Metrics for 2020 Disclosure Avoidance},
    author={{U.C. Census Bureau}},
    year={2020},
    month={March},
    publisher={U.S. Census Bureau},
    howpublished={Available at: \url{https://www2.census.gov/programs-surveys/decennial/2020/program-management/data-product-planning/disclosure-avoidance-system/2020-03-25-data-metrics-2020-da.pdf} (Accessed August 2022)}
}

@misc{hawesspence,
    title={Understanding the 2020 Census Disclosure Avoidance System: Production settings and DAS accuracy metrics for the
P.L. 94-171 Redistricting Data Summary File},
    year={2021},
    month={July},
    author={Michael Hawes and Matthew Spence},
    publisher={U.S. Census Bureau},
    howpublished={Available at: \url{https://www2.census.gov/about/training-workshops/2021/2021-07-01-das-presentation.pdf}}
}

@misc{apportionment,
    title={Why Is the U.S. Census So Important?},
    author={Mark Mather and Paola Scommegna},
    publisher ={Population Reference Bureau},
    year={2019},
    month={March},
    note={Available at: \url{https://www.prb.org/resources/importance-of-u-s-census/} (Accessed August 2022)}
}

@misc{flaxmanrepo,
  author = {Abraham Flaxman and Beatrix Haddock},
  title = {Simulation Study of Re-identification Risk in {ppmf\_20210428\_eps12-2\_P}},
  publisher = {GitHub},
  journal = {GitHub repository},
  howpublished = {Available at: \url{https://github.com/aflaxman/ppmf_12.2_reid} (Accessed June 2022)},
}

@jurisdiction{baldrige,
  title = {Baldrige v. Shapiro},
  volume = {455},
  reporter = {U.S.},
  pages = {345},
  date = {1982},
}

@misc{censustract,
  title={DATA GEM: What is a Census Tract? Making Sense of Census Geography},
  note={\\ \url{https://www.census.gov/data/academy/data-gems/2018/tract.html}},
  author={{U.S. Census Bureau}},
  publisher={U.S. Census Bureau}
}

@misc{tractsize,
  title={Census Tracts for the 2020 Census-Final Criteria},
  note={\url{https://www.federalregister.gov/documents/2018/11/13/2018-24567/census-tracts-for-the-2020-census-final-criteria}},
  publisher={U.S. Census Bureau},
  date={November 13, 2018}
}
\end{document}